\documentclass[prd,twocolumn,superscriptaddress,nofootinbib,colorlinks=true]{revtex4-1}
\usepackage{amsmath, amssymb, graphicx, hyperref, slashed}
\usepackage[dvipsnames]{xcolor}

\pdfoutput=1

\def\lag{\mathcal{L}}

\def\beq{\begin{equation}}
\def\eeq{\end{equation}}

\def\bit{\begin{itemize}}
\def\eit{\end{itemize}}
\def\ben{\begin{enumerate}}
\def\een{\end{enumerate}}

\def\lag{\mathcal{L}}
\def\half{{1 \over 2}}
\def\sphi{\varphi}
\def\Mplanck{M_\text{Pl}}
\def\Msolar{M_\odot}
\def\MeV{\,\text{MeV}}
\def\GeV{\,\text{GeV}}
\def\TeV{\,\text{TeV}}
\def\cm{\,\text{cm}}
\def\km{\,\text{km}}

\def\Eq#1{Eq.~\ref{#1}}

\begin{document}

\title{Asymmetric Dark Stars and Neutron Star Stability}

\author{Moira I. Gresham}
\affiliation{Whitman College, Walla Walla, WA 99362}
\author{Kathryn M. Zurek}
\affiliation{Theory Group, Lawrence Berkeley National Laboratory, Berkeley, CA 94720}
\affiliation{Berkeley Center for Theoretical Physics, University of California, Berkeley, CA 94720}
\affiliation{Theory Group, CERN, Geneva, Switzerland}

\begin{abstract} 
We consider gravitationally bound states of asymmetric dark matter (ADM stars), and the impact of ADM capture on the stability of neutron stars.  We derive and interpret the equation of state for ADM with both attractive and repulsive interactions, and solve the Tolman-Oppenheimer-Volkoff equations to find equilibrium sequences and maximum masses of ADM stars. Gravitational wave searches can utilize our solutions to model exotic compact objects (ECOs). 
Our results for attractive interactions differ substantially from those in the literature, where fermionic ADM with attractive self-interactions was employed to destabilize neutron stars more effectively than non-interacting fermionic ADM. 
By contrast, we argue that fermionic ADM with an attractive force is no more effective in destabilizing neutron stars than fermionic ADM with no self-interactions.  
\end{abstract}

\maketitle


\section{ Introduction}  Work on hidden sector dark matter has exploded over the last decade \cite{Battaglieri:2017aum}.  One conclusion of this work is that even modest extensions of the standard paradigm of dark matter---as a single, stable, weakly interacting particle coupling only via Standard Model forces---to include the dynamics of dark forces can easily change the cosmology, astrophysics and terrestrial signatures of the dark sector \cite{Alexander:2016aln}.  
A leading example of this is theories of asymmetric dark matter (ADM), where coupling to dark forces arises naturally as a means to annihilate the symmetric abundance of dark matter \cite{Kaplan:2009ag,Lin:2011gj,Zurek:2013wia,Petraki:2013wwa}, similar to the annihilation of electron-positron pairs to photons in the early Universe.  When dark forces are present, the cosmology of DM is generically modified due to self-interactions \cite{Spergel:1999mh,Loeb:2010gj,Lin:2011gj,Tulin:2013teo,Tulin:2012wi,Kaplinghat:2015aga,Tulin:2017ara}. 

When dark ADM forces are sufficiently strong and attractive, bound states can form, similar to the formation of nuclei in the Standard Model \cite{Wise:2014jva,Wise:2014ola,Gresham:2017zqi}.  If the dark sector simultaneously lacks a repulsive long range force (the analogue of the photon), very large states---nuggets---are generically synthesized \cite{Wise:2014jva,Hardy:2014mqa,Krnjaic:2014xza,Gresham:2017cvl,Gresham:2018anj}.

The same dynamics that leads to nugget formation in the early Universe can also lead to the formation of ADM stars in the late Universe, via condensation arising from radiation of dark force mediators or small nugget fragments \cite{Gresham:2018anj}. Self-interacting or not, ADM may also collect in neutron stars.  If the ADM is a scalar particle, a black hole can form, destroying the parent neutron star  
 \cite{Goldman:1989nd, Kouvaris:2010jy, Kouvaris:2011fi, McDermott:2011jp, Kouvaris:2012dz, Guver:2012ba, Bell:2013xk, Bertoni:2013bsa}.  If the ADM is a fermion, Fermi degeneracy pressure tends to stabilize the ADM, though in principle attractive self-interactions can help to overcome degeneracy pressure.

The primary results in this paper are a self-consistent set of mass-radius relationships and stability bounds of exotic compact objects (ECOs) comprised of fermionic ADM with attractive and/or repulsive self-interactions.  We also consider the impact of such fermionic ADM on neutron star stability. 
We focus on a model with a single stable Dirac spin-1/2 fermion, $X$, as the dark matter candidate, with attractive self-interactions mediated by a real scalar, $\phi$, and repulsive self-interactions by a vector, $V^\mu$:\footnote{The choice of the signs on $g_\phi$ and $g_V$ make $\langle \phi \rangle$ and $\langle V^0 \rangle$ positive when $g_V$ and $g_\phi$ are positive.}
\begin{multline}
\lag = \bar X \left[ i \slashed{\partial } - g_V \slashed V - (m_X- g_\phi \phi) \right] X \\ -{1 \over 4} V_{\mu \nu}^2 + {1 \over 2} m_V^2 V_\mu^2
+ \half (\partial_\mu \phi \partial^\mu \phi - m_\phi^2 \phi^2) - V(\phi). \label{eq: lagrangian}
\end{multline}
We solve the Tolman-Oppenheimer-Volkoff (TOV) equations for this theory to determine the stability against gravitational collapse.
We will argue that for ECOs composed of fermionic constituents with arbitrary self-interactions, the smallest maximum stable ECO size for a given mass scale per constituent, $m_X$, is approximately realized in this model when only a scalar mediator with negligible potential is present. Then we will see that this minimum is of the same order of magnitude as Landau's estimate for Fermi degeneracy supported matter: $N_\text{max} \sim \Mplanck^3/m_X^3, M_\text{max} \sim \Mplanck^3/m_X^2$ \cite{1932PhyZS...1..285L}.


We find in particular that spin-1/2 ADM with an attractive force never collapses neutron stars (NSs) over their lifetime in our Universe unless (perhaps) the fermionic constituents are heavier than order $10^6 \GeV$.  This means that fermionic ADM with an attractive force does not in general solve the missing pulsar problem \cite{Kouvaris:2011gb,Bramante:2014zca,Bramante:2015dfa},  
 have limits from imploding NSs \cite{Bramante:2013nma}, or lead to non-primordial solar mass black holes \cite{Kouvaris:2018wnh}. 
 We also find a different equilibrium sequence for stars made of self-attractive ADM than derived elsewhere \cite{Kouvaris:2015rea}.  The most important difference between our work and previous treatments is use of a fully consistent equation of state (EoS), instead of utilizing a Yukawa potential valid only in the non-relativistic and low-density limit in the case of scalar-mediated interactions.\footnote{The EoS derived from the Yukawa potential is identical to the fully relativistic EoS in the vector-mediated case but not in the scalar-mediated case.} This crucially changes both the impact on NS stability 
 and the ADM star equilibrium sequence. Our fully relativistic treatment extends to a fully general relativistic treatment of ECOs composed of two possibly interacting but separately conserved constituents---in our case baryonic matter and fermionic ADM.

The potential structure and stability of fermionic dark matter ECOs \cite{Narain:2006kx,Kouvaris:2015rea} and of ADM-admixed NSs \cite{Sandin:2008db,Ciarcelluti:2010ji,Leung:2011zz,Xiang:2013xwa,Zheng:2014fya,Zheng:2016ygg,Mukhopadhyay:2015xhs} has been examined before, and there has been renewed interest in such objects  in the context of gravitational wave observations \cite{Giudice:2016zpa,Nelson:2018xtr,Kopp:2018jom,Das:2018frc, Abbott:2018oah}. This work 
gives a comprehensive account of the effect of interactions, including the first correct treatment of attractive self-interactions that cause binding, and baryon-ADM interactions in the case of admixed stars. 


The outline of this paper is as follows. In Sec.~\ref{sec: equation of state} we specify and interpret the EoS for spin-1/2 dark matter with attractive and/or repulsive scalar- and/or vector-mediated self-interactions. In Sec.~\ref{sec: ADM stars} we find and interpret the sequence of gravitationally stable stars composed of such matter. 
Then in Sec.~\ref{sec: collapse}, employing results from the previous section, we argue that ADM with spin-1/2 constituents smaller than about $10^6 \GeV$ cannot collapse NSs, regardless of whether the constituents are self-interacting.  Finally, in Sec.~\ref{sec: conclusion} we conclude. Appendix \ref{app: two sequences} explains non-generic features of ADM at the cusp of being self-bound. Appendix \ref{app: admixed} lays out the general relativistic equations appropriate for determining structure and gravitational stability of static stars composed of multiple separately conserved, possibly interacting, components. It details methods we used to obtain numerical solutions that confirm the less technical arguments presented in Sec.~\ref{sec: collapse}.

\section{Equation of State for Self-interacting Spin-1/2 ADM}\label{sec: equation of state}
The EoS for fermionic matter described by \Eq{eq: lagrangian} is given by \cite{1955PhRv...98..783J,1956PhRv..103..469D,1974AnPhy..83..491W,Walecka:1995mi,Glendenning:1997wn}
\begin{align}
\epsilon &= {m_X^4 \over 3 \pi^2} \bigg(\frac{\sphi^2}{2C_\phi^2} + W(\sphi) + C_V^2 \frac{(k_F/m_X)^6}{2} \nonumber \\
& \qquad \qquad + 3 \int_0^{k_F/m_X}  {x^2}{\sqrt{x^2 + (1-\sphi)^2}} \, dx \bigg), \label{eq: energy density} \\
p &={m_X^4 \over 3 \pi^2} \bigg(- {\sphi^2  \over {2 C_\phi^2}} - W(\sphi)  + C_V^2 \frac{(k_F/m_X)^6}{2} \nonumber \\
& \qquad \qquad +  \int_0^{k_F/m_X} { x^4 \over \sqrt{x^2 + (1-\sphi)^2}} dx \bigg), \label{eq: pressure}
\end{align}
where
\beq
C_i^2 \equiv {4 \alpha_i \over 3 \pi}{m_X^2 \over m_\phi^2}, \label{eq: C squared}
\eeq
with $\alpha_i \equiv g_i^2/4 \pi$, $k_F$ the $X$ Fermi momentum, $W(\sphi) \equiv {m_X^4 \over 3 \pi^2} V(m_X \sphi / g_\phi)$, and $\sphi=g_\phi \langle \phi \rangle / m_X$ is defined through the transcendental equation
\beq
{\sphi \over C_\phi^2} + W'(\sphi) = 3 \int_0^{k_F/m_X} x^2 {1 - \sphi \over \sqrt{x^2 + (1 - \sphi)^2}} dx. \label{eq: mstar}
\eeq
The effective Dirac mass (c.f.~\Eq{eq: lagrangian}) is $m_* = m_X(1-\sphi)$ and the number density is given by 
$
n = 2 \int^{k_F} {d^3 \vec k \over (2 \pi)^3}= {k_F^3 \over 3 \pi^2}. \label{eq: number density}
$
The equations above assume zero temperature, though generalization to nonzero temperature is straightforward and has been worked out in the context of the $\sigma$-$\omega$ model (see {\em e.g.}~\cite{Walecka:1995mi,Glendenning:1997wn}). The equations are derived in the mean field limit, where scalar and vector fields are approximated by their mean values. Additionally the mean fields are assumed to be static and spatially uniform. This last assumption is inconsistent with solutions to the general relativistic equilibrium equations when order one variations in star density occur over length scales comparable to or smaller than the force range, $1/m_\phi, 1/m_V$. For example with $C_\phi^2$ fixed, the approximation breaks down in the decoupling limit, $\alpha_\phi \rightarrow 0$. We explored the transition where spatial uniformity breaks down for self-bound matter in \cite{Gresham:2017zqi}.

Eqs.~\ref{eq: energy density} and \ref{eq: pressure} are related through the thermodynamic relation
\beq
p = - {\partial E \over \partial V}\big|_{S,N} = - {\partial (\epsilon / n) \over \partial (1 / n)} = {\partial \epsilon \over \partial n} n - \epsilon = \mu n - \epsilon \, ,  \label{eq: thermo relation}
\eeq
where $\mu = {\partial \epsilon \over \partial n}$ is chemical potential. Note that rest energy per constituent, $\epsilon/n$, is necessarily minimized when $p=0$. For large enough attractive interactions, there are solutions where $p=0$, ${\partial p \over \partial n} > 0$, and the binding energy per particle is positive ($m_X - \epsilon/n > 0$), meaning that large stable self-bound states exist, elsewhere called nuggets \cite{Wise:2014ola,Gresham:2017zqi,Gresham:2017cvl,Gresham:2018anj}.

\begin{figure}
\includegraphics[scale=0.85]{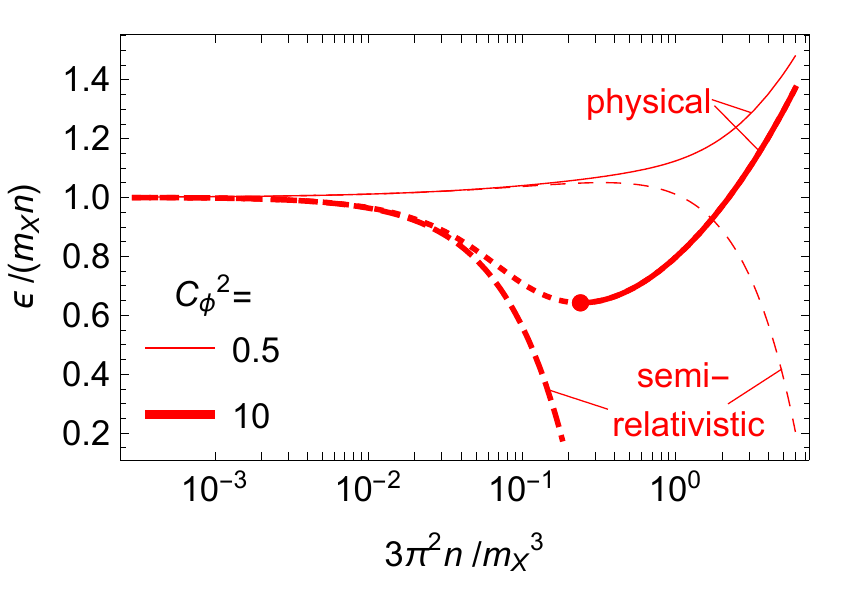}
\caption{Energy per particle per $m_X$ as a function of number density for ${C_\phi^2} \equiv {4 \alpha_\phi \over 3 \pi} {m_X^2 \over m_\phi^2} = 0.5$ (thin) and $10$ (thick) using a semi-relativistic treatment (dashed) alongside the expressions obtained using relativistic mean field theory (solid).  When $C_\phi^2=10$, ADM is self-bound; the dotted line shows the analytic EoS in the density range below the density of self-bound nuggets, where the matter is unstable to coalescence.}\label{fig: EoS}
\end{figure}

Fig.~\ref{fig: EoS} shows the rest energy per particle as a function of number density for ${C_\phi^2} =0.5, 10$, $C_V^2=0$, and $V(\phi)=0$. The solid/dotted lines show \Eq{eq: energy density} while the ``semi-relativistic'' dashed lines show the energy computed assuming $\epsilon = \epsilon_\text{kin}+\epsilon_Y$ with potential energy given by the nonrelativistic expression $\epsilon_Y =- \half n^2 \alpha_\phi \int \int {e^{-m_\phi r_{i j}} \over r_{i j}} d^3 \vec{r}_i d^3 \vec{r}_j / \text{Volume}$, with indices $i$ and $j$ running over particles, and $\epsilon_\text{kin} = 2 \int^{k_F} \sqrt{k^2+m_X^2} {d^3 \vec{k} \over (2 \pi)^3}$, as was done explicitly or implicitly in Refs.~\cite{Kouvaris:2011gb,Bramante:2013nma,Bramante:2014zca,Kouvaris:2015rea,Bramante:2015dfa,Kouvaris:2018wnh}, for example. As number density grows and the constituents grow more relativistic, the fully relativistic and semi-relativistic expressions diverge. Were it correct, the semi-relativistic expression would imply that self-attractive ADM is microscopically unstable such that the energy per constituent becomes negative at high density and is unbounded below. By contrast, the fully relativistic expression for energy per constituent remains positive but can develop a local or global minimum at nonzero density.  This happens because Fermi pressure overcomes the attractive force at high density and the pressure grows again. More specifically, as density grows, $\langle \phi \rangle$ grows, initially decreasing pressure, but simultaneously decreasing the effective Dirac mass, accelerating the growth of Fermi degeneracy pressure.

For $ C_\phi^2 > 1.09$ there is a global minimum in $\epsilon/n$, with energy per constituent at this minimum less than $m_X$ (the value at $n=0$), indicating the existence of large stable bound states, or nuggets,\footnote{For large enough $C_\phi^2$, local and global minima also exist when $C_V^2\neq0$, $V(\phi) \neq 0$. See e.g.~\cite{Gresham:2017zqi},\cite{Gresham:2018anj}.} 
 that form in principle without the aid of gravity. The analog of surface tension causes large states to form spheres \cite{Gresham:2017zqi}. The large point on the $C_\phi^2=10$ curve in Fig.~\ref{fig: EoS} represents the saturation density---the density of large self-bound nuggets. The dotted curve lies in a density region with mostly negative pressure, representing the instability of ADM to condensation into large, dense nuggets.  
 
Numerical solutions to Eqs.~\ref{eq: energy density} and \ref{eq: mstar} for $0.840 < C_\phi^2 < 1.09$ reveal a local minimum with $\epsilon/n > m_X$, indicating a phase change at positive pressure. At this pressure, zero-temperature matter jumps from a gas-like state at low density to a liquid-like state at higher density. The liquid state can be realized only with the help of another force---for example with the aid of gravity in the core of a star. See Appendix~\ref{sec: maxwell} for further discussion.

With the EoS for ADM self-interacting through scalar and vector exchange in hand, we now explore the structure of self-gravitating objects composed of such matter.

\section{ADM Star Stability and Equilibrium Sequence}\label{sec: ADM stars}

In the previous section, we showed that self-attractive fermionic ADM is microscopically stable. Our main objective here is to pinpoint when cold ADM stars become gravitationally unstable. In particular we will show that the maximum stable mass cannot deviate much below Landau's estimate, $M_\text{max}^\text{fermion} \sim \Mplanck^3/m_X^2$ \cite{1932PhyZS...1..285L} (see also \cite{1965gtgc.book.....H}).  An implication is that self-attractive fermionic ADM cannot seed collapse of neutron stars more efficiently than non-self-interacting fermionic ADM.

We restrict our attention to spherically symmetric compact objects such that the Tolman-Oppenheimer-Volkoff (TOV) equation governing such objects reads,
\beq
{d p \over d r} = - {(p + \epsilon) (G M(r)/r + 4 \pi G r^2 p) \over r(1-2 G M(r)/r) } \label{tov}
\eeq  or equivalently
\beq
c_s^2 {d \ln n \over d \ln r} = - {(G M(r)/r + 4 \pi G r^2 p) \over (1-2 G M(r)/r) } \label{tov}
\eeq
where $r$ is the radial coordinate, $M(r) = 4 \pi \int_0^r \epsilon \, r'^2 dr'$, $c_s^2 = {d p \over d \epsilon}$ is squared sound speed that characterizes the stiffness of matter, and we have used \Eq{eq: thermo relation}. 
Given an EoS relating $\epsilon$ and $p$, this single integro-differential equation can be solved for any given choice of central energy density by integrating out from $r = 0$ to the edge of the star $r=R$ where $p(R)=0$. Eqs.~\ref{eq: energy density} and \ref{eq: pressure} represent a parametric EoS, $\left\{ p(n), \epsilon(n) \right\}$, which leads to an integro-differential equation for number density, $n$, as a function of $r$. The number of constituent particles in the star is given by $N_X = \int n(r)/\sqrt{1-2 G M(r)/r} \, d^3 \vec{r}$.

\begin{figure*}
\includegraphics[width=0.6 \textwidth]{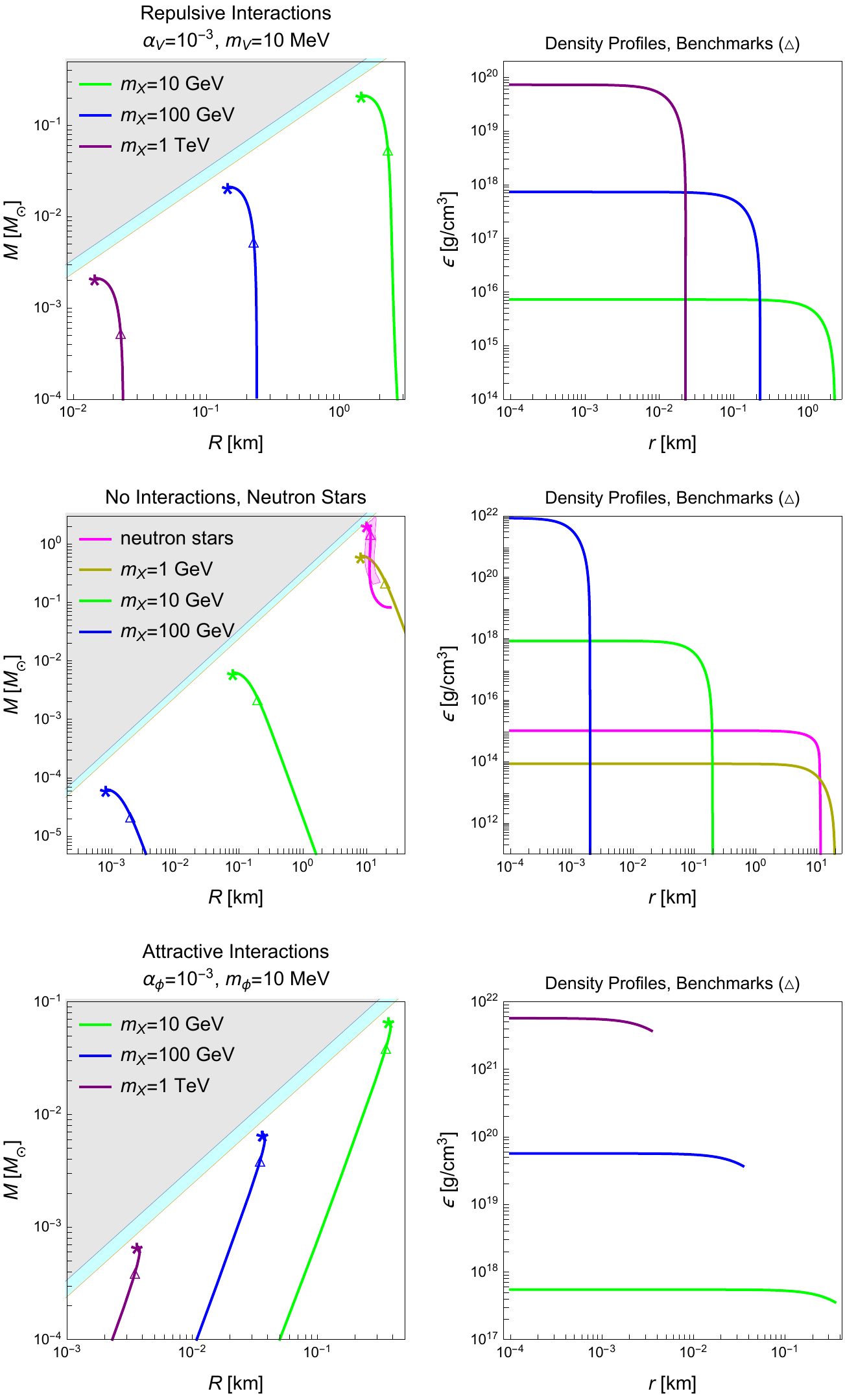}
\caption{Equilibrium sequences for varying $m_X$ represented by star mass as a function of radius for spin-1/2 matter with repulsive (top), no (middle),  and attractive (bottom) self-interactions with fixed mediator mass and coupling. See also~Table~\ref{tab: max mass formulas}.   Asterisks mark the stable equilibrium sequence endpoints corresponding to the global maximum in $M$ as a function of central density. The gray region corresponds to $R < R_s = 2 G M$ and the cyan contour represents maximum compactness, $G M / R  =  0.354$. Compare Fig.~3 in \cite{Kouvaris:2015rea}; the repulsive case agrees but the attractive case dramatically differs due to differences in the EoS. Right-hand plots show energy density for a benchmark star (marked with $\triangle$ in the left-hand plots) as a function of distance from its center.  The cutoffs at finite density in the bottom right-hand plot indicate the discontinuity in energy density at $r=R$ due to self-boundedness.  For comparison, the middle plot also shows the equilibrium sequence for a sample NS matter EoS (magenta) consistent with NS observations to date---the HB EoS as defined in \cite{Read:2009yp}. The NS benchmark is a $1.5 \Msolar$ star, and the shaded magenta region is digitized from \cite{Annala:2017llu}.  
}\label{fig: mX varies equilibrium sequences}
\end{figure*}

\begin{figure*}
\includegraphics[width=0.6 \textwidth]{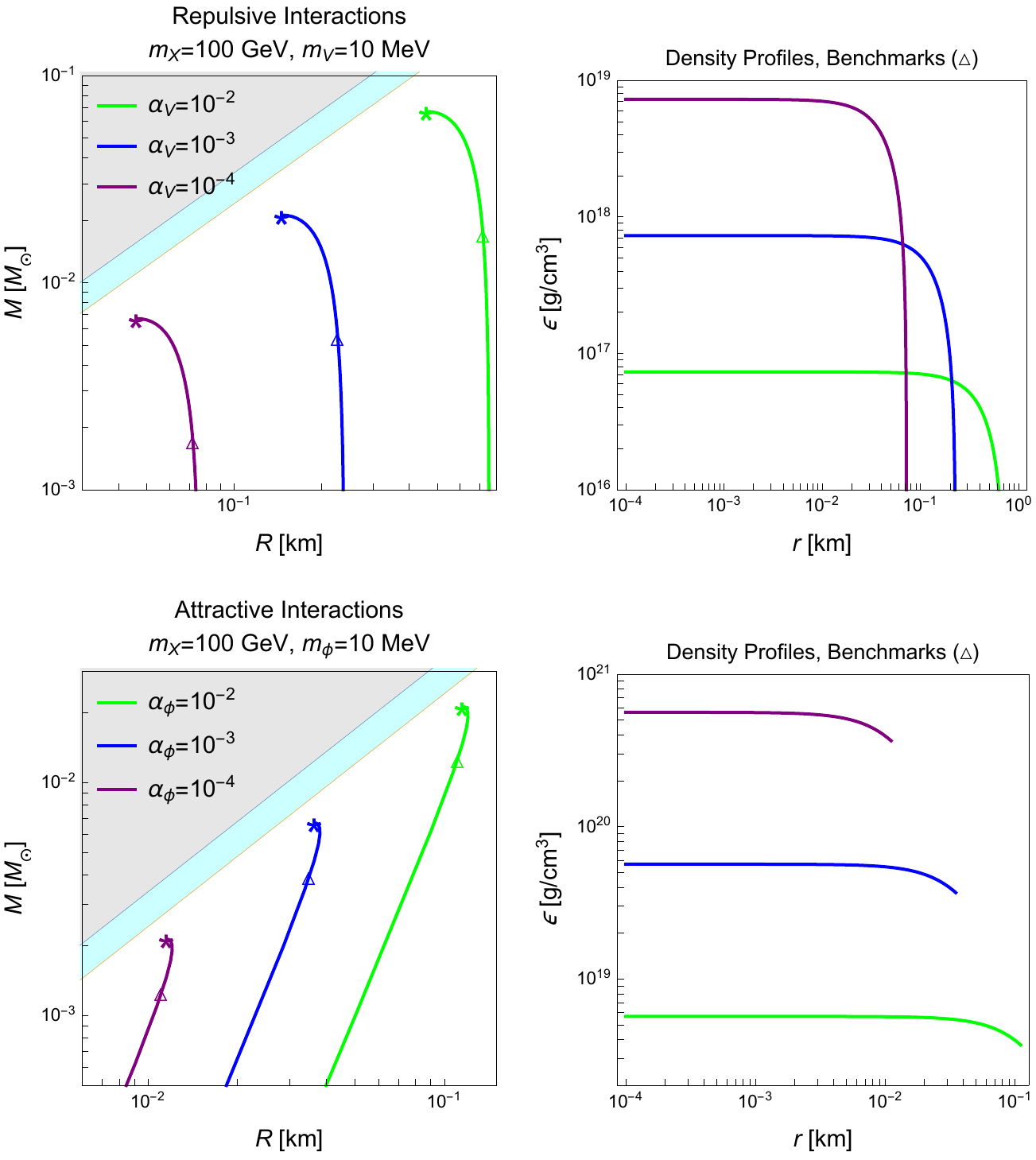}
\caption{Equilibrium sequences for varying $\alpha$ represented by star mass as a function of radius for spin-1/2 matter with repulsive (top) or attractive (bottom) self-interactions with fixed mediator and constituent mass.  Compared to Fig.~4 in \cite{Kouvaris:2015rea}, the repulsive cases agree but the attractive cases disagree. See Fig.~\ref{fig: mX varies equilibrium sequences} for further detail. }\label{fig: alpha varies equilibrium sequences}
\end{figure*}

\begin{figure*}
\includegraphics[scale=1]{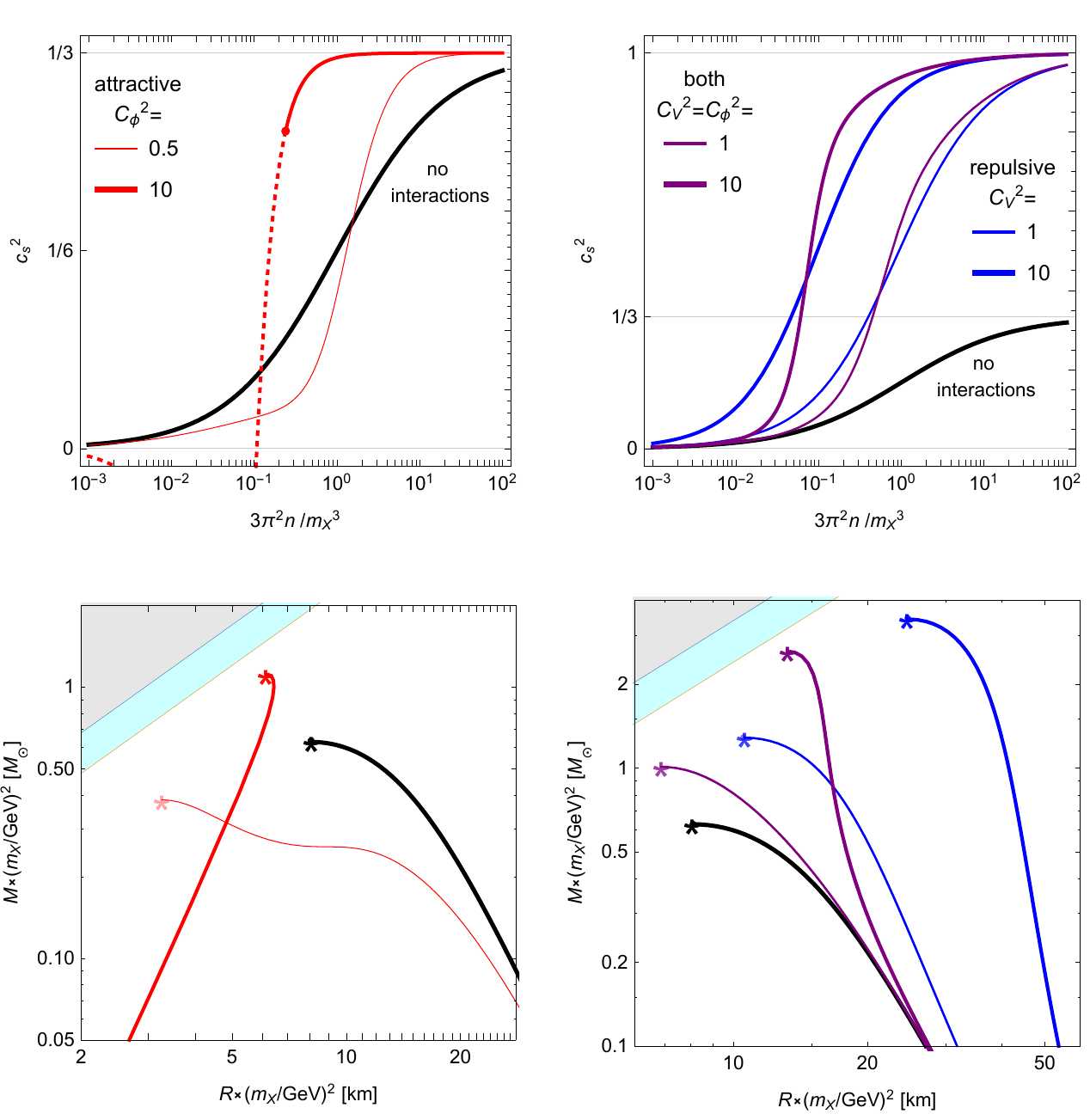}\\ 
\caption{ \emph{Top:} Squared sound speed, $c_s^2 = {d p \over d \epsilon}$, as a function of $(k_F/m_X)^3 $  for spin-1/2 dark matter with attractive (red), repulsive (blue), both attractive and repulsive (purple), and no (thick black) self-interactions. The strength of attractive and repulsive self-interactions is characterized by $C_\phi^2$ and $C_V^2$, respectively, as defined in \Eq{eq: C squared}. Here $k_F$ is Fermi momentum and $X$ number density is $n = {k_F^3 / 3 \pi^2}$.  The dotted section of the red $c_s^2$ curve corresponds to the dotted region in Fig.~\ref{fig: EoS}, where matter is unstable to condensation into large self-bound states with density marked by the dot. \emph{Bottom:} Mass and radius of static, spherically symmetric stars composed of such matter, representing the equilibrium sequence. The curves cut off at the maximum-mass gravitationally stable stars, denoted with asterisks. Gray regions correspond to ($R < R_\text{s} = 2 G M$). The cyan boundary marks ${G M \over R} = 0.354$, corresponding to the theoretically most compact non-black-hole objects \cite{1997ApJ...488..799K,Lattimer:2015nhk}.}\label{fig: equilibrium sequences}
\end{figure*}

The gravitational stability of ADM stars is calculated from the equilibrium configurations, which are solutions to the TOV equations for a given central density.  Maxima in mass as a function of central density indicate a transition from stable to unstable  \cite{1965gtgc.book.....H}. 
Figs.~\ref{fig: mX varies equilibrium sequences}, \ref{fig: alpha varies equilibrium sequences}, and \ref{fig: equilibrium sequences} show the mass and radius of solutions to the TOV equations for spin-$1/2$ ADM with various self-interaction strengths. 
Figs.~\ref{fig: mX varies equilibrium sequences} and \ref{fig: alpha varies equilibrium sequences}  show similar results to those in \cite{Narain:2006kx,Kouvaris:2015rea} for repulsive interactions.  However, for $m_X\sim \{10 \GeV,100 \GeV,1 \TeV \} $ the attractive interactions corresponding to $m_\phi = 10 \MeV, \alpha_\phi \in \{10^{-2},10^{-3},10^{-4}\}$ \emph{increase} the maximum mass (and maximum number of constituents, $N_X$) of a stable gravitationally bound ECO relative to the case of non-interacting fermions of the same mass, in contrast to the results in \cite{Kouvaris:2015rea}. For comparison, the magenta region in Fig.~\ref{fig: mX varies equilibrium sequences} includes equilibrium configurations for NS matter that can support a $2 M_\odot$ star, satisfy the 90\% confidence level constraint on tidal deformability from the NS binary inspiral gravitational wave observation, GW170817 \cite{TheLIGOScientific:2017qsa, Abbott:2018exr}, and are consistent with known limits on the baryonic matter EoS at (low) nuclear densities and at very high densities \cite{Annala:2017llu}.   

The attractive self-interaction parameters shown in Figs.~\ref{fig: mX varies equilibrium sequences} and \ref{fig: alpha varies equilibrium sequences} all correspond to $  C_\phi^2 = {4 \alpha_\phi \over 3 \pi}{m_X^2 \over m_\phi^2} > 100$ and thus to the case of strong self-binding; the growth of radius with mass until near the gravitational stability endpoint (here marked with asterisks) is characteristic of compact objects made of self-bound matter such as those of hypothetical self-bound strange quark matter stars \cite{Haensel:1986qb}. 
The increased gravitational stability of these objects relative to their non-interacting counterparts stems from their effectively stiffer EoS due to enhanced Fermi degeneracy pressure all the way out to the edge of the star. For self-interactions satisfying $C_\phi^2 \gg 1$ and negligible scalar potential, we have identified universal formulas for $N_\text{max}, M_\text{max}$ and $\left({G M \over R}\right)_\text{max}$. We report these in Table~\ref{tab: max mass formulas} alongside analogous relations for the cases of no interactions, purely repulsive interactions, and attractive-repulsive interactions with $C_V^2=C_\phi^2$.\footnote{Our formula for $M_\text{max}$ in the purely repulsive case matches that reported in \cite{Narain:2006kx}.} The formulae work well when $C_i^2 \gtrsim 10$. In the attractive case, the asymptotic values may alternatively be written $N^\text{attractive}_\text{max} = 0.34 \Mplanck^3/\bar m_X^3$, $M^\text{attractive}_\text{max} = 0.28 \Mplanck^3/\bar m_X^2$ where $\bar m_X$ is the zero-pressure chemical potential, equivalent to the average mass per constituent of large nuggets. For $C_\phi^2 \gg 1$, $\bar m_X \rightarrow (2/C_\phi^2)^{1/4}$ \cite{Gresham:2017zqi,Gresham:2018anj}. An alternative explanation of the enhanced gravitational stability of matter with large  attractive self-interactions is that these interactions decrease the effective constituent mass scale, so the Landau limit $M_\text{max} \sim \Mplanck^3/m^2$ still applies but with $m$ to be interpreted as $\bar m_X$ where $\bar m_X < m_X$.

A scalar potential term of the form $\lambda  \phi^4$ with $\lambda > 0$ tends to limit $\bar m_X$ from below \cite{Gresham:2017zqi, Gresham:2018anj}, reducing the possibility of substantially raising $M_\text{max}$ for fixed $m_X$. In general, such a term tends to push $c_s^2$ closer to its form with no interactions, indicating that scalar potentials tend to push the maximum mass and other equilibrium sequence characteristics closer to that for no interactions.

\renewcommand{\arraystretch}{1.5}
\begin{table*}
\begin{tabular}{| c | | c | c | c | c |}
\hline
& no interactions & attractive &  both  & repulsive \\
& $C_V^2=C_\phi^2=0$ & $C_V^2=0, C_\phi^2 \gg 1$ & $C_V^2=C_\phi^2 \gg 1$ & $C_V^2 \gg 1, C_\phi^2=0$ \\
\hline
\hline
$N_\text{max}$ 	& $0.399  \, {\Mplanck^3 /  m_X^3} $ 	& $0.34 \left({C_\phi^2 / 2}\right)^{3/4} {\Mplanck^3 /  m_X^3}$ 	& $0.61 \sqrt{C_V^2} \, {\Mplanck^3 /  m_X^3}$	& $0.69 \sqrt{C_V^2} \, {\Mplanck^3 /  m_X^3}$ \\
$M_\text{max}$ 	&  $0.384 {\Mplanck^3 /  m_X^2}$ 		& $0.28  \left({C_\phi^2 / 2}\right)^{2/4} {\Mplanck^3 /  m_X^2}$				& $0.47 \sqrt{C_V^2} \, {\Mplanck^3 /  m_X^2}$   & $0.63 \sqrt{C_V^2} \, {\Mplanck^3 /  m_X^2}$ \\
$\left({G M \over R}\right)_\text{max}$	& 0.115 & 0.27 & 0.35  & 0.21\\
\hline
\end{tabular}
\caption{Mass, $M$, number of constituents, $N$, and compactness, ${G M \over R}$, for the static spherically symmetric maximum-mass stars made of spin-1/2 matter with large attractive, repulsive, both attractive and repulsive, and no interactions. The dimensionless constants $C_i^2 = {4 \alpha_i \over 3 \pi}{m_X^2 \over m_i^2}$ characterize interaction strength. In the attractive case, the matter is strongly self-bound with the chemical potential at zero pressure equal to $\mu|_{p=0} = \bar m_X = m_X (2/C_\phi^2)^{1/4}$. In all other cases shown the matter is not self-bound so $\mu|_{p=0} = m_X$. For quick reference, note: $\Mplanck^3 = 1.63 \Msolar \GeV^2$.  }\label{tab: max mass formulas}
\end{table*}

Fig.~\ref{fig: equilibrium sequences} shows equilibrium sequences for matter with more moderate ($C_\phi^2, C_V^2 \lesssim 10$) attractive, repulsive, or both attractive and repulsive self-interactions side-by-side with squared sound speed, $c_s^2 = {d p \over d \epsilon}$, characterizing the stiffness of the matter. 
The figure demonstrates that \emph{softer equations of state lead to smaller maximum masses and vice versa}: the larger the density range where $c_s^2 $ lies below the no-interactions curve, the smaller the maximum mass relative to the no-interactions case and vice versa. In the top left plot, we see that any softening of the EoS relative to the no-interactions EoS accelerates the approach to the high density limit, $c_s^2 \sim 1/3$. And in the top right figure, comparing purple to blue, we again see that attractive interactions soften the EoS at low densities but stiffen it a larger densities, accelerating the approach to the high density limit when a vector is present, $c_s^2 \sim 1$.  In all examples, the more extreme the softening at lower densities, the more extreme the stiffening at higher densities. This can also be seen analytically as follows.

The chemical potential
for the model \Eq{eq: lagrangian} is given by 
\begin{align} 
{\mu } & = C_V^2 {k_F^3 \over m_X^2} + \sqrt{k_F^2 + m_*^2(k_F)}, \label{eq: chem pot} 
\end{align}
with $m_*(k_F)$ determined through \Eq{eq: mstar}.
With a scalar potential guaranteeing positive energy density and therefore microscopic stability, one can show that $m_*$  approaches zero in the large-density limit; and larger $C_\phi^2$ drives $m_*$ to zero faster while a quartic potential term moderates the decrease.  For spin-$1/2$ matter in general, $c_s^2 = {dp \over d \epsilon} = {1 \over 3}{d \ln \mu \over d \ln k_F}$ by \Eq{eq: thermo relation} and $n={k_F^3/3 \pi^2}$. In the large density limit, vector repulsion dominates pressure and $\mu \sim k_F^3/m_X^2$ if a vector is present, or fermi pressure dominates and $\mu \sim k_F$ if the vector is absent; thus  $c_s^2 \rightarrow 1~\text{or}~1/3 $  with or without vector repulsion, respectively. Furthermore since attractive interactions cause $m_*$ to decrease, though this initially drives $c_s^2$ below its value absent the attractive interactions, it also accelerates the approach to the asymptotic limit. This explains the tendency of attractive interactions to soften the matter at lower densities and stiffen it at higher densities.

The left plots in Fig.~\ref{fig: equilibrium sequences} represent matter with attractive self-interactions, including examples of non-self-bound matter ($C_\phi^2=0.5$) and self-bound matter ($C_\phi^2=10$). The equilibrium sequence for $C_\phi^2=0.5$ begins at low central density and large radius following the no-interactions sequence, but ends at lower maximum mass and smaller radius. By contrast, since self-bound matter fuses before reaching a cool state, the $C_\phi^2 = 10$ equilibrium sequence begins at relatively large densities and small radii, and never meets up with the no-interactions sequence. 
For $C_\phi^2 < 0.516$ and $C_\phi^2 > 1.09$, the first maximum in $M$ as a function of central density is the global maximum.

In the purely attractive, $V(\phi)=0$ case, the equilibrium solutions for $0.516 < C_\phi^2 < 1.09$ hug the no-interactions sequence at low central density and large radius, and then develop a local maximum at low compactness before reaching the global maximum at larger compactness, indicating the existence of two separate sequences analogous to the white dwarf and NS sequences (see e.g.~\cite{Glendenning:1997wn,Balberg:2000xu}).  We further discuss this range in Appendix~\ref{sec: two sequences}, but note that the first local maximum occurs in the range $ 0.04 \Mplanck^3/m_X^2 \lesssim M_\text{max}^\text{local} \lesssim 0.15 \Mplanck^3/m_X^2$---less than a factor of ten lower than the no-interactions global maximum of $0.38 \Mplanck^3/m_X^2$---except in the narrow range $1.05 < C_\phi^2 < 1.09$. 

For purely attractive interactions and fixed $m_X$, the global maxima in the entire $C_\phi^2$ range satisfy $M_\text{max} > 0.23 \Mplanck^3 / m_X^2$,  $N_\text{max} > 0.24 \Mplanck^3 / m_X^3$, and $(G M / R)_\text{max} > 0.092$ with the bounds saturated when $C_\phi^2=0.45$, $C_\phi^2=0.42$, and $C_\phi^2=0.26$, respectively. In each case, the parameter decreases from the no interactions value to the value at the minimum, and then increases monotonically toward the asymptotic value in Table~\ref{tab: max mass formulas}. 

In the case of equal strength attractive and repulsive interactions, as the interactions become more extreme with $C_\phi^2 = C_V^2 \gg 1$, the speed of sound curve approaches a step function, $c_s^2 \sim \theta(n - n_\text{crit})$. Matter with such behavior is thought to produce the theoretically most compact stars, and indeed we find that $\left({ G M \over R}\right)_\text{max} \sim 0.354$, the posited maximum in the literature assuming subluminal sound speed \cite{1997ApJ...488..799K,Lattimer:2015nhk}, for $C_\phi^2 = C_V^2 \ggg 1$. (See also Table \ref{tab: max mass formulas}.)

We now consider self-interacting fermionic ADM more generally. If the cost of softening matter at a given density through attractive interactions is accelerating the approach to the asymptotic limit, generally, then the softness of microscopically stable fermionic matter is limited, and therefore the amount that self-interactions can reduce the maximum stable mass below that for free fermionic matter is limited. On this basis, we conjecture that $M_\text{max} \gtrsim 0.1 \Mplanck^3 / m_X^2$, $N_\text{max} \gtrsim 0.1 \Mplanck^3 / m_X^3 $ holds true for spin-1/2 ADM, generally.

\section{Implications for Neutron Star Collapse}\label{sec: collapse}
So far we have focused on ADM-only stars, including pinpointing the maximum mass of gravitationally stable self-attractive ADM stars.  Based on a semi-relativistic calculation, Refs.~\cite{Kouvaris:2011gb,Bramante:2013nma,Bramante:2014zca,Bramante:2015dfa,Kouvaris:2018wnh} have claimed that ADM with an attractive force, captured inside of NSs, can destabilize and destroy them.  Here we argue that relativistic effects stabilize the NS over most of the parameter space, and destabilization can occur only for very large ADM mass of order PeV.

\subsection{ADM Capture}\label{sec: capture}

The amount of ADM captured in a NS in a time $t$ is at most the amount that impacts the NS \cite{Goldman:1989nd},
\beq
 m_X {N_X}_\text{cap} \lesssim  \int  \langle \pi b_\text{max}^2 \rho_\text{DM} v_\text{DM} \rangle dt \label{MDM captured 1}
\eeq
where $b_\text{max} = \sqrt{{2 G M/R \over 1-2 G M/R}} {R \over v_\text{DM}}$ is the impact parameter corresponding to DM that just scrapes the surface of the NS at closest approach.\footnote{If the ADM is already bound as nuggets at the time of capture, then replace $m_X$ with $\bar m_X$ in \Eq{MDM captured 1} and \Eq{MDM captured}.} Here the energy density $\rho_\text{DM}$ and velocity scale $v_\text{DM}$ are to be taken asymptotically far away from the NS. For typical NSs, $G M / R \sim 0.2$ and $R \sim 10 \text{km}$ leading to 
\beq
m_X {N_X}_\text{cap} \lesssim \left( 3 \times 10^{-14} \Msolar \right) {\rho_\text{DM} \over \GeV/\cm^3}{200 \text{km/s} \over v_\text{DM}}{t \over 10^{10} \text{yrs}}. \label{MDM captured}
\eeq
The upper bound is realized only when on average 100\% of the DM passing through the NS deposits enough energy to be captured. For DM with mass of order GeV to PeV, the minimum required baryon-DM cross section for this to occur is order $2 \times 10^{-45} \text{ cm}^2$ \cite{Goldman:1989nd}. For $m_X \lesssim \text{GeV}$, Fermi blocking suppresses the scattering \cite{Bertoni:2013bsa}, and for $m_X \gtrsim \text{PeV}$, multiple scatters are required for gravitational capture \cite{Bramante:2017xlb} and therefore greater cross sections are required to realize the upper bound in \Eq{MDM captured}. If the ADM is bound in large composite states at the time of capture, additional considerations apply \cite{Hardy:2015boa}. In general since the density and velocity scales entering \Eq{MDM captured} are not vastly different from the fiducial values, the amount of ADM captured is a tiny fraction of the total mass of a NS (order $1.5 \Msolar$). One can speculate about other ways to realize ADM-admixed NSs stars with a much larger fraction of ADM than can be collected gradually through capture over the NS lifetime. One interesting possibility is copious production and capture of dark matter in the core-collapse supernova of the NS's progenitor \cite{Nelson:2018xtr}.  

\subsection{ADM-Admixed Neutron Stars}\label{sec: admixed}

For ADM captured by a NS over its lifetime to induce its collapse, a self-gravitating ADM star of the same mass as the captured ADM must itself be unstable to collapse.  Based on this observation, we argue that capture of spin-1/2 ADM---self-interacting or not---with constituent masses smaller than order $10^6 \GeV$ by NSs cannot in any circumstances induce collapse.  

As detailed in Appendix~\ref{app: admixed}, we solved the general relativistic equilibrium equations for NS matter admixed with cold ADM as modeled in Sec.~\ref{sec: equation of state}. 
In general, the maximum number of ADM constituents, $({{N_X}|_{N_b \neq 0}})_\text{max}$, possible in a stable ADM-admixed NS with fixed baryon number can be smaller than the maximum number of ADM constituents in an ADM-only star $({{N_X}|_{N_b = 0}})_\text{max}$ made of the same kind of ADM. However, we find that any appreciable differences between $({{N_X}|_{N_b \neq 0}})_\text{max}$ and $({{N_X}|_{N_b = 0}})_\text{max}$ occur only if $m_X ({{N_X}|_{N_b = 0}})_\text{max}$ is comparable to a solar mass. 
Including baryon-ADM interactions does not affect this conclusion.  And including thermal effects only increases stability. 

Now assume the amount of captured ADM is a small fraction,
\beq
f \equiv (m_X {N_X}_\text{cap}) / M_\text{NS} \ll 1,
\eeq
of the NS mass, as predicted by \Eq{MDM captured}, and that the NS is not already teetering at its stability bound with mass greater than $2 \Msolar$. Then as long as an ADM-only star of size $N_X = {N_X}_\text{cap}$ does not collapse, neither will an ADM-admixed NS with the same amount and type of ADM. From our treatment of fermionic ADM-only stars, since 
\beq
 m_X {N_X}_\text{max}^\text{ADM-only} \gtrsim 0.1 \Mplanck^3/m_X^2,
\eeq 
taking $M_\text{NS} \sim 1.5 \Msolar$ we find
\beq
{
m_{X,\,\text{collapse}} \gtrsim  
(10 \, f)^{-1/2}  \GeV .} \label{m collapse}
\eeq 
Assuming maximally efficient ADM capture over 10 billion years and galactic ADM densities $\rho_\text{DM} \sim 1$-$100 \GeV/\text{cm}^3$ and the velocity scale $v_\text{DM} \sim 200 \km/\text{s}$, we find
\beq
m_{X,\,\text{collapse}} \gtrsim 2 \times 10^5 \,\text{-}\, 2 \times 10^6 \GeV.
\eeq We have not yet argued that ADM can collapse NSs, but rather only that ADM with spin-1/2 constituent mass $m_X <  (10\, f)^{-1/2} \GeV$ cannot collapse NSs if $f \ll 1$.

Now we consider whether ADM \emph{can} collapse NSs in certain cases when $m_X >  (10\, f)^{-1/2} \GeV$. More specifically, if an ADM-only star with constituents $N_X = {N_X}_\text{cap}$ is unstable to gravitational collapse, then is an  ADM-admixed NS with $N_X = {N_X}_\text{cap}$ also  unstable?  If there is a process to cool the ADM sufficiently that it self-gravitates, the answer is yes. 
Let us briefly consider the ADM cooling process after capture. Ref.~\cite{Baryakhtar:2017dbj} estimates the cross section needed to maximize ADM capture, saturating \Eq{MDM captured}, and the time required for the ADM to deposit most of its kinetic energy in the NS. When $m_X \gtrsim 10^6 \GeV$ and \Eq{MDM captured} is saturated, for example, a captured ADM particle deposits most of its kinetic energy in less than a day. The ADM heats the NS, which will be detectable by the next generation of infrared telescopes \cite{Baryakhtar:2017dbj}. Again if the ADM-baryon cross section is anywhere near the level required to maximize ADM capture, then according to the estimates in \cite{McDermott:2011jp}, thermalization of the ADM with baryonic matter happens quickly. Accounting for NS heating through ADM capture, the maximum blackbody temperature of NSs near Earth is expected to be $1750\, \text{K}$ ($0.15 \,\text{eV}$) \cite{Baryakhtar:2017dbj}, which is a minuscule fraction of the Fermi momentum of near-collapse ADM with $m_X\gtrsim10^6 \GeV$. The ADM indeed cools to effectively zero temperature and for ADM with $M^\text{ADM-only}_\text{max} \ll \Msolar$, once the amount of captured ADM approaches this maximum, the ADM core density within the NS far exceeds the baryon density. The ADM core self-gravitates and its structure is unaffected by the baryonic matter. When ${N_X}_\text{cap} > {N_\text{max}^\text{ADM-only}}$, the ADM core is unstable to collapse.     

\section{Conclusions and Outlook}\label{sec: conclusion}
We have argued that the capture of spin-$1/2$ ADM by neutron stars cannot lead to their implosion unless  the mass of the spin-$1/2$ ADM constituents exceeds approximately one PeV. This includes ADM with attractive or other varieties of self-interactions. Thus the existence of old NSs can only set limits on ADM-baryon cross sections for spin-1/2 ADM constituent masses larger than about one PeV. Once the next generation of infrared telescopes comes online, limits from dark kinetic heating of NSs in this mass range may compete with any such limits \cite{Baryakhtar:2017dbj, Bell:2018pkk}. If there is a positive detection of dark kinetic heating of NSs, the existence of old neutron stars will provide complementary information on possible models of ADM with $m_X \gtrsim \text{PeV}$.

After deriving and interpreting the equation of state for cold spin-1/2 ADM self-interacting through scalar and/or vector mediators, we found solutions to the general relativistic gravitational equilibrium equations for cold static spherically symmetric ADM stars and identified the maximum size of gravitationally stable ADM stars. We found formulas for this maximum stable size in the case of strong self-interactions (see Table~\ref{tab: max mass formulas}). We also found that the maximum size at fixed $m_X$ generically does not drop below $M^\text{star}_\text{max} = 0.1 \Mplanck^2/m_X^2$, and we conjectured that a similar limit holds for spin-1/2 ADM with arbitrary self-interactions. 

If fermionic ADM stars are realized in our Universe, they might be detectable by gravitational wave observatories \cite{Aasi:2013wya, TheLIGOScientific:2014jea} in the event of mergers with other compact objects \cite{Giudice:2016zpa}. The masses and radii of these objects can be drastically different from those of NSs, though not necessarily so.  As compared to a NS binary merger, we expect the electromagnetic signature of a merger involving at least one ADM star to be a smoking gun signal of the difference even if the gravitational wave form does not reveal the presence of the ECO. As compared to a black hole binary merger, the waveform will be modified due to the ADM star's spatial structure and tidal deformability.  We leave the prospects of gravitational wave detection of ADM stars for future work.


\section*{Acknowledgments}

We thank Hou Keong (Tim) Lou for prior collaboration and Daniel Ega\~{n}a for comments on a draft.  MG is supported by the NSF under award No.~1719780. KZ is supported by the DoE under contract No.~DE-AC02-05CH11231.


\appendix


\section{Phase Change and Secondary Equilibrium Sequences}\label{app: two sequences}

Purely self-attractive ADM with $C_\phi^2 > 1.09$ is self-bound, not relying on gravity for its boundedness. Here we focus on ADM that is not quite self-bound.  

Zero temperature matter with purely attractive interactions of strength $0.840 < C_\phi^2 < 1.09$ and $V(\phi)=0$, as modeled in Sec.~\ref{sec: equation of state}, undergoes a phase change at positive pressure and number density.  Further, when $0.516 < C_\phi^2 < 1.09$, the equilibrium sequence obtained by solving the TOV equations has an additional unstable region as compared to the sequences described in the main text.  Rest energy per constituent, sound speed, and gravitational equilibrium mass and density configurations are shown for this range of couplings in Fig.~\ref{fig: two sequences}.  The dotted regions show \emph{unphysical} solutions to Eqs.~\ref{eq: energy density}-\ref{eq: mstar}, signaled by a local minimum in energy per constituent as a function of number density and a negative squared sound speed, $c_s^2 = {d p \over d \epsilon}$. The dashed regions in the bottom plots represent unphysical solutions to the TOV equations, separating two separate gravitationally stable equilibrium sequences that emerge due to a temporary stall in the growth of sound speed with density. In the next two subsections we discuss how the {\em physical} solutions are constructed, first examining the equation of state, and then considering the equilibrium sequence obtained from the TOV equations.

\subsection{Phase Change and Maxwell's Construction for the Equation of State}\label{sec: maxwell}

We first consider the rest energy per constituent and speed of sound, shown in the upper two panels of Fig.~\ref{fig: two sequences}.   The local minimum in the analytic $\epsilon / n$,  $C_\phi^2=1$ curve at nonzero density indicates a phase change. Matter lying near the local maximum in $\epsilon/n$ can lower its energy per constituent (and pressure) by condensing, and matter between the local maximum and minimum has negative pressure. As seen in the upper right panel of Fig.~\ref{fig: two sequences}, sound speed also becomes imaginary near the local maximum. These features all signal the unphysical nature of the analytic EoS in this density domain.
The physical EoS is obtained by choosing endpoints ${n}_A$ and ${n}_B$ that lie at equal pressure on either side of the density region with $\partial p / \partial n < 0$ such that 
$
\left( \partial \epsilon \over \partial {n} \right)_A = \left( \partial \epsilon \over \partial {n} \right)_B = {\epsilon_B - \epsilon_A \over {n}_B - {n}_A}. 
$ 
This construction is equivalent to Maxwell's construction in standard thermodynamics \cite{1965gtgc.book.....H}, and is shown by the solid red line labeled ``physical'' in Fig.~\ref{fig: two sequences}.    Matter at densities less than ${n}_A$ exists in a stable gas-like state and matter at densities greater than ${n}_B$ exists in a stable liquid-like state. When there is a phase change, we use ``liquid'' and ``gas''  to refer to the high-density and low-density phases, respectively. Matter in the density region between ${n}_A$ and ${n}_B$ coexists in two different density states (liquid and gas). The phase change occurs at nonzero pressure; to realize this nonzero pressure, some other force must be applied---for example, gravity. The liquid state can exist in the cores of gravitationally bound stars; in static spherically symmetric stars an abrupt transition from liquid core to gas crust occurs at a given radius.

\subsection{Two Equilibrium Sequences when $0.516 < C_\phi^2 < 1.09$}\label{sec: two sequences}

\begin{figure*}
\includegraphics[scale=1]{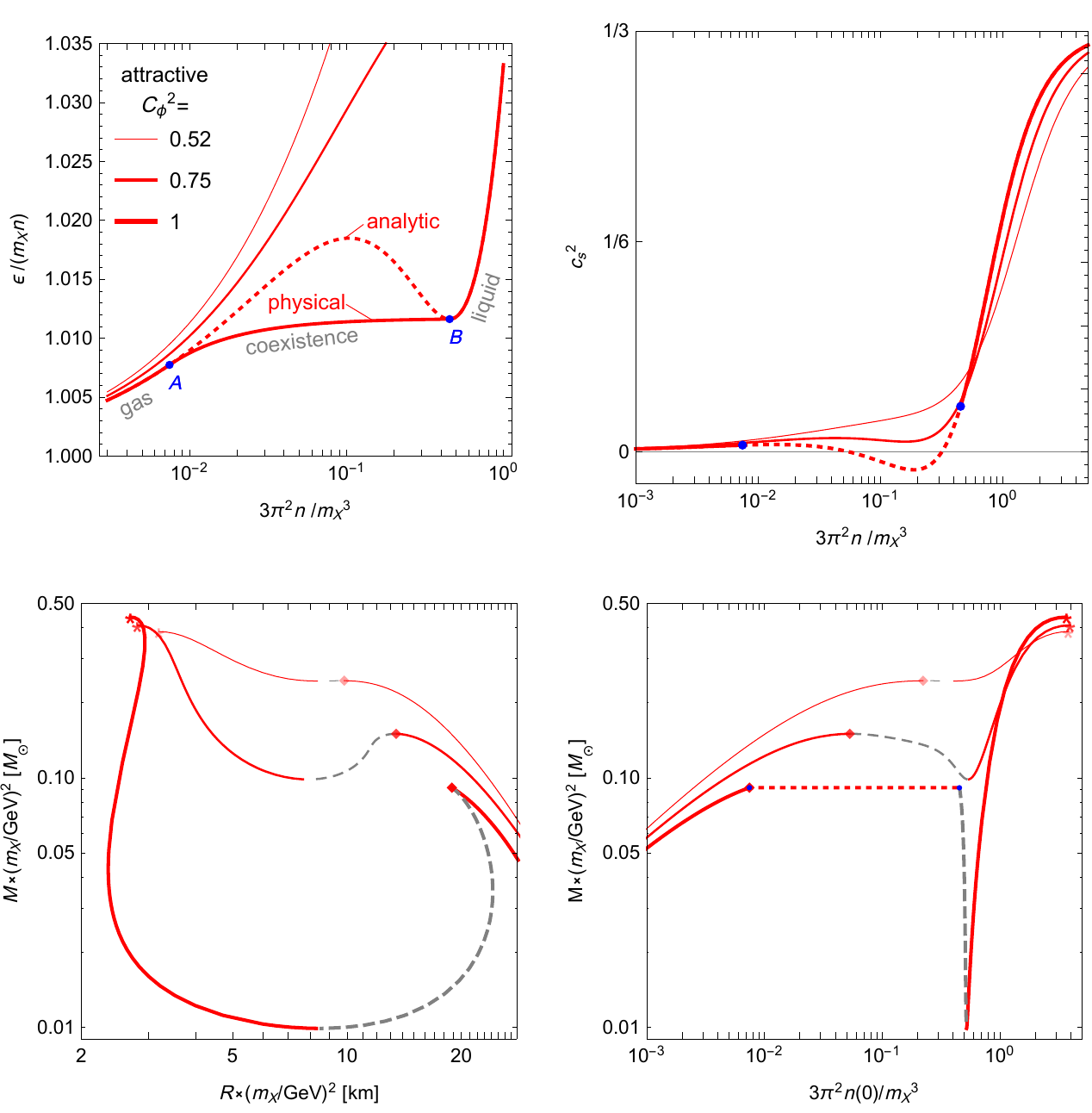}
\caption{EoS as represented by energy per constituent per mass, $\epsilon / (m_X n)$ (top left), and sound speed squared, $c_s^2 = {d p \over d \epsilon}$ (top right) as functions of density, alongside solutions to the TOV equations as represented by star mass versus star radius (bottom left) or versus central number density, $n(0)$ (bottom right) for matter with attractive interaction strengths $C_\phi^2=0.52, 0.75, 1$. Blue dots on the $C_\phi^2=1$ curve mark the matching points for Maxwell's construction; the matter is in its liquid phase at densities $n > n_B$, in its gas phase at densities $n < n_A$, and coexisting in both phases in between. The dotted red lines in the top panels represent the unphysical analytic EoS in the coexistence density range.  In a star, the phases do not coexist but rather density abruptly jumps from $n_A$ to $n_B$ at a given radius, and equilibrium solutions with central density $n_A < n(0) < n_B$ do not exist; this region is marked with a dotted red line in the bottom right panel, which maps onto the cusp in the bottom left panel. The asterisks mark the maximum mass stable star in the higher-density NS-like sequence while diamonds mark that of the lower-density white dwarf-like sequence. Dashed gray lines indicate gravitationally unstable equilibrium solutions between the two sequences. See the text for further detail.  
}\label{fig: two sequences}
\end{figure*}

Next we consider solutions to the TOV equations.  For purely attractive interactions and $0.516 < C_\phi^2 < 1.09$, the sequence of solutions to the TOV equations contains a local maximum in $M$ as a function of central density  before hitting a global maximum, as shown in the bottom right panel of Fig.~\ref{fig: two sequences}. The local maximum at lower density occurs at larger radius and lower compactness, as shown in the bottom left panel. This same behavior occurs for cold catalyzed matter due to a relative softening of the EoS near the neutron drip density; near this density the speed of sound temporarily decreases (softening) before increasing again (stiffening) as a function of density (see e.g.~\cite{1965gtgc.book.....H,Glendenning:1997wn,Balberg:2000xu}). Correspondingly, a local maximum and minimum in star mass as a function of central density develops near the neutron drip density, matching onto the stability endpoint of the white dwarf sequence and the beginning of the NS sequence, respectively. In Fig.~\ref{fig: two sequences}, analogs of the white dwarf sequence are represented by solid lines stretching from low central density and larger radius up to the local maxima marked by diamonds. Analogs of the NS sequence are represented by solid lines that stretch from the local minima in mass to the global maxima at higher central densities and smaller radii, marked by asterisks.

Another way of characterizing the feature of cold catalyzed matter that leads to the separate white dwarf and NS sequences is that its sound speed growth (or stiffening) temporarily \emph{stalls} near the neutron drip density. This is similar to what we see in the top right panel of the figure. For $C_\phi^2=0.52$, the stall is moderate, with $c_s^2$ continuing to increase but at a lower rate near densities $n \sim 0.2 {m_X^3 \over 3 \pi^2}$.  Correspondingly, as seen in the bottom right panel, the growth of gravitational mass with central density near $n(0) \sim 0.2 {m_X^3 \over 3 \pi^2}$ stalls so much that a local maximum develops, signaling an instability to contraction. The dashed gray region lying between local maximum and minimum represents gravitationally unstable solutions to the TOV equations. In this case, the instability is relatively mild, with stability taking hold again at only slightly higher central densities.  For $C_\phi^2=0.75$, the stall in sound speed growth is more severe. Correspondingly, the the minimum in star mass as a function of central density is deeper, and the density gap between equilibrium sequences is larger. The stall grows more severe with increasing $C_\phi^2$.

As demonstrated by the $C_\phi^2=1$ curve of the bottom left panel, for matter with a phase change, a cusp develops near the local maximum in the TOV solutions for $M(R)$ and is associated with the discontinuity in density because of the phase change. Solutions to the TOV equations with central densities corresponding to the coexistence density range do not exist; there is a gap in the mass versus central density curve where $n_A < n(0) < n_B$, marked by the dotted red line in the lower right panel.  This entire gap is mapped onto the cusp in the lower left panel.  Such behavior also occurs in models of compact stars that include QCD phase changes, see {\em e.g.}~\cite{ROSENHAUER1992630,Alford:2017qgh}. 

Understanding the final states of stars or ADM cores within NSs that reach the white dwarf-like stability endpoint---be they NS-like or black holes---is beyond the scope of this work. But we remark that the white dwarf-like mass endpoints lie within a factor of ten below the no-interactions global stability endpoint except in the range very near the transition to self-bound matter, $1.05 < C_\phi^2 < 1.09$. 

\section{ADM-Admixed Neutron Stars and Baryon-ADM Interactions}\label{app: admixed}

In the bulk of the paper we focused on ADM-only stars.  Here we lay out the equations for baryon-ADM admixed stars. Then based on our solutions to these equations, we argue that, unless the mass of the ADM and baryons in the star is of the same order, the amount of ADM that destabilizes a NS is little modified from the amount that destabilizes an ADM-only star.

\subsection{Gravitational Equilibrium Equations}
To investigate ADM-admixed NSs, we need the analog of the TOV equations for two interacting but separately conserved matter species. The TOV equations are equivalent to extremizing mass, 
\beq
M = \int \epsilon(r)\, 4 \pi {r}^2 dr, \label{eq: M definition}
\eeq  with baryon number, 
\beq N_b = \int n_b(r) {4 \pi {r}^2 dr \over \sqrt{1 - 2 G M(r)/r}}, \label{eq: N definition}
\eeq held fixed \cite{1965gtgc.book.....H}. To generalize to multiple conserved species, we extremize $M$ with each conserved species number separately held fixed through the method of Lagrange multipliers. That is, for ADM-baryon stars, extremize the functional $F=M-\mu_b N_b - \mu_X N_X$, treating baryon density, $n_b$, and ADM density, $n_X$, as independent functions. The Lagrange multipliers $\mu_i$ are gravity-inclusive chemical potentials.   
The derivation is worked out for $N$ such species in detail in Ref.~\cite{1972PThPh..47..444K}.
One finds,
\beq
\mu_i = e^{\nu(r)} {\partial \epsilon \over \partial n_i} = \text{constant} \label{chem potential}
\eeq with
\beq
{d \nu \over d r} = { G M(r) / r + 4 \pi G r^2 p \over r (1-2 G M(r)/r)} \label{metric TOV}
\eeq
where $\mu_i$ is the gravity-inclusive chemical potential of species $i$, $p = \sum_i {\partial \epsilon \over \partial n_i} n_i - \epsilon $, and $\nu$ is a metric function defined by $|g_{t t}| = e^{2 \nu}$ . The TOV equation, \Eq{tov}, is equivalent to $\sum_i n_i {d \over d r} \mu_i = 0$, with $\nu$ eliminated through \Eq{metric TOV}.

The total mass $M$, constituent numbers $N_X, N_b$, and radius $R$, are determined by the equilibrium equations, Eqs.~\ref{chem potential}-\ref{metric TOV}, along with the equations defining $M$ and $N_i$, Eqs.~\ref{eq: M definition}-\ref{eq: N definition}, for given central baryon and ADM number densities, $\{ {n_b}(0), {n_X}(0) \}$.\footnote{We will discuss a caveat when significant baryon-ADM interactions are present.} 
 We interpret the largest stable star at fixed baryon number, ${N_b}_0$, to correspond to the first local maximum in $M$ as a function of $\{ {n_b}(0), {n_X}(0) \}$, subject to the constraint $N_b = {N_b}_0$. Since equilibrium configurations satisfy $dM = \mu_b dN_b + \mu_X dN_X = 0$, and since $\mu_i$ are finite for any equilibrium configuration, with fixed $N_b={N_b}_0$, $M$ and $N_X$  attain their maxima at the same $\{ {n_b}(0), {n_X}(0) \}$.\footnote{We have generalized from the observation of Landau. See \cite{1965gtgc.book.....H}.} When the ADM constitutes a tiny fraction of the star by mass, it is numerically easier to identify the maximum in $N_X$.

Absent DM-baryon interactions, the number density of a given species affects the other only through the total gravitational mass and pressure. In this case, \Eq{chem potential} and \Eq{metric TOV} are equivalent to  
\beq
{d p_i \over d r} = - {(p_i + \epsilon_i)\left( G M(r)/r + 4 \pi r^2 G p \right) \over r(1-2 G M(r)/r)}, \label{general tov 2}
\eeq where $p_i = n_i {\partial \epsilon_i \over \partial n_i} - \epsilon_i$. Given a large hierarchy between the densities of the two species, as we will detail below, even a weak DM-baryon interaction can dramatically affect the density profile of the subdominant species where the two species overlap; in this case it is important to use Eqs.~\ref{chem potential}-\ref{metric TOV} rather than \Eq{general tov 2}. 
Furthermore, in such cases, we find there can be multiple distinct equilibrium configurations corresponding to zero central density of the component with lighter constituents. 
In the next section we describe how to include baryon-ADM interactions before discussing our numerical solutions to the gravitational equilibrium equations with and without ADM-baryon interactions in Sec.~\ref{sec: numerical solutions}.

\subsection{Modeling Baryon-ADM Interactions}\label{sec: baryon-adm interactions}

Using relativistic mean field theory and the same techniques used in the context of the $\sigma$-$\omega$ model of nuclear physics (see e.g.~\cite{Walecka:1995mi,Glendenning:1997wn}), we find that a vector-mediated baryon-ADM interaction gives rise to an interaction energy density and pressure
\beq
\epsilon_I = p_I = {g_b g_X \over m_{A'}^2} n_b n_X, \label{eq: interaction energy density}
\eeq where $g_b$ and $g_X$ are the vector-nucleon and vector-ADM coupling constants, respectively, and $m_{A'}$ is the vector mediator mass. The low energy elastic nucleon-$X$ scattering section is given by  $\sigma_{b X} = {\mu_{b X}^2 \over \pi} \left({g_b g_X \over m_{A'}^2}\right)^2$ with $\mu_{b X}$ the reduced mass, so \Eq{eq: interaction energy density} is alternatively written,
\beq
\epsilon_I = p_I = \sqrt {\pi \sigma_{b X} } \left( {m_X + m_b \over m_X m_b} \right) {n_b} {n_X },
\eeq
where $m_b \approx  \GeV$ is the nucleon mass scale. The total energy density in an admixed star is $\epsilon = \epsilon_b + \epsilon_X + \epsilon_I$ with $\epsilon_b$ independent of $n_X$, and $\epsilon_X$ independent of $n_b$, and similarly for pressure.\footnote{This clean decomposition of energy density and pressure does not occur for scalar mediators of ADM-baryon interactions.} 

Consider $m_X \gg \GeV$. Current direct detection constraints on $\sigma_{b X}$ in this range are $\sigma_{b X} \lesssim 10^{-46} \left({m_X \over 100 \GeV }\right) \text{cm}^2 \approx 10^{-19} \left({m_X \over 100 \GeV }\right) \GeV^{-2} $ \cite{Akerib:2016vxi,Cui:2017nnn,Aprile:2018dbl}. Baryon number densities (energy densities) toward the centers of NSs are order $10^{-2} \GeV^3(\GeV^4)$. Thus given interaction strengths near current direct detection limits, interaction energy density is comparable to baryon energy density when $\sqrt{m_X \over \GeV} n_X \gtrsim 10^{10} \GeV^3 $.
The number density of free fermionic ADM reaches order $m_X^3$ in the cores of near-collapse stars, implying the bound can be satisfied for $m_X \gtrsim \TeV$ dark matter.  

The hierarchy $\epsilon_I, \epsilon_b \ll \epsilon_X$ with $\epsilon_I \gtrsim \epsilon_b$ naturally occurs for $m_X  \gtrsim \TeV $ when ADM-baryon interactions are near current direct detection limits. In this case, both the gravitational and non-gravitational ADM-baryon interactions affect the baryon density profile in the small admixed core, while the dark matter density profile is unaffected by the baryonic matter.

Conversely, direct detection bounds on sub-GeV dark matter become weak, and furthermore for ADM masses much smaller than a $\GeV$, we expect $X$ number (energy) densities to be much smaller than order $\GeV^3~(\GeV^4)$. 
In this case ADM-baryon interactions can be important in determining density profiles of sub-GeV dark matter within ADM-admixed NSs.

\subsection{Results: Numerical Solutions to the Equilibrium Equations}\label{sec: numerical solutions}

\begin{figure*}
\hspace{0.2 in} Free ADM, $m_X = 100\, m_b$ \hspace{2.3 in}   Free ADM, $m_X = m_b$ \\
\includegraphics[width=0.95\textwidth]{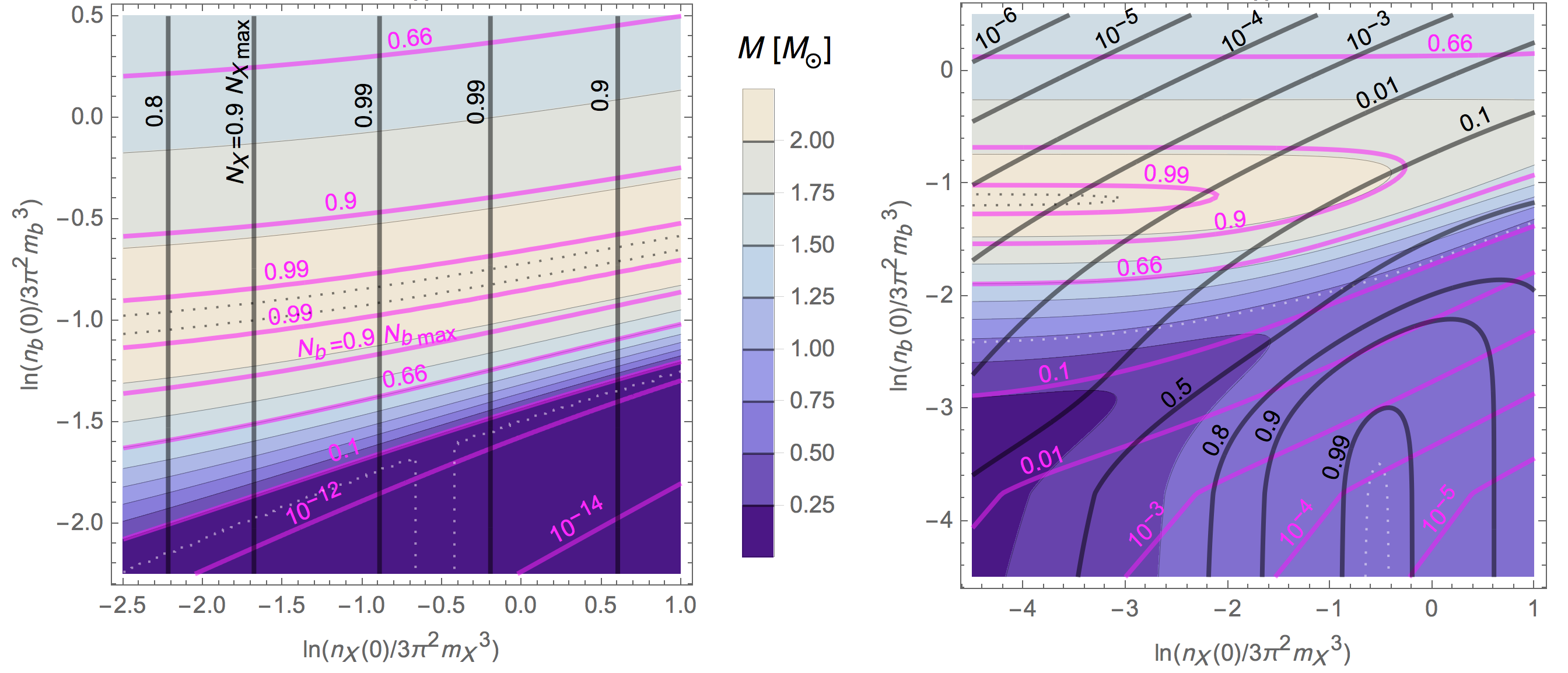}
\caption{Contours of constant mass $M$ in units of $M_\odot$ (multicolor shading), ADM number $N_X$ as a fraction of $N_{X\,\text{max}}^\text{ADM-only}=0.399 (\Mplanck/m_X)^3$ (black), and baryon number $N_b$ as a fraction of  $N_{b \, \text{max}}^\text{baryons-only} = 1.37 (\Mplanck/m_b)^3$  (magenta) for solutions to the gravitational equilibrium equations, as functions of central ADM and baryon number densities, with $m_X=100\, m_b$ (\emph{left}) or $m_X=m_b$ (\emph{right}) and $m_b=939.5 \MeV $ (\emph{both}). These plots demonstrate that, unless the amount of ADM is of the same order or larger than the baryonic matter by mass ($m_X N_X \gtrsim m_b N_b$), the amount of ADM that destabilizes a NS is little modified from the amount that destabilizes an ADM-only star. 
Configurations lying below the maximum in $N_b$ (at fixed $N_X$) and to the left of the maximum in $N_X$ (at fixed $N_b$) are stable.  Other configurations are gravitationally unstable. The ADM is modeled as a free fermi gas and the baryonic matter through a spliced polytrope, HB as in Fig.~\ref{fig: mX varies equilibrium sequences}. Note that the maximum mass of a baryons-only star for the same model is $2.12 \Msolar$ (for reference the dotted black line is the $M=2.115 \Msolar$ contour). For free spin-1/2 ADM-only stars, the maximum mass is $0.627 (\GeV/m_X)^2 \Msolar$ (for reference by the dotted white line is the $M=0.626 (\GeV/m_X)^2 \Msolar$ contour). For  $m_X > 100 \, m_b$, the plot is unchanged relative to that for $m_X = 100 \, m_b$. In the entire region of the left-hand plot, $m_X N_X \lll m_b N_b$.  Stable equilibrium configurations for any choice of $N_X$ and $N_b$ exist as long as they are smaller than the maximum value for the corresponding single-component stars. 
By contrast, $m_X N_X \sim m_b N_b \sim \Msolar$ toward the middle of the right-hand plot, where we see that stable equilibrium configurations with both $N_X$ and $N_b$ near their maxima for single-component stars do not exist. However, when $m_X N_X \ll m_b N_b \sim \Msolar$ (toward the top left of the plot) the maximum mass of $2.12 \Msolar$ is unaffected.  In either case, configurations with any $m_X N_X \lll m_b N_b < m_b {N_b}^\text{baryons-only}_\text{max}$ exist as long as $N_X < {N_X}^\text{ADM-only}_\text{max}$. }\label{fig: free ADM admixed NS contour plot}
\end{figure*}

\begin{figure*}
\includegraphics[width=0.9\textwidth]{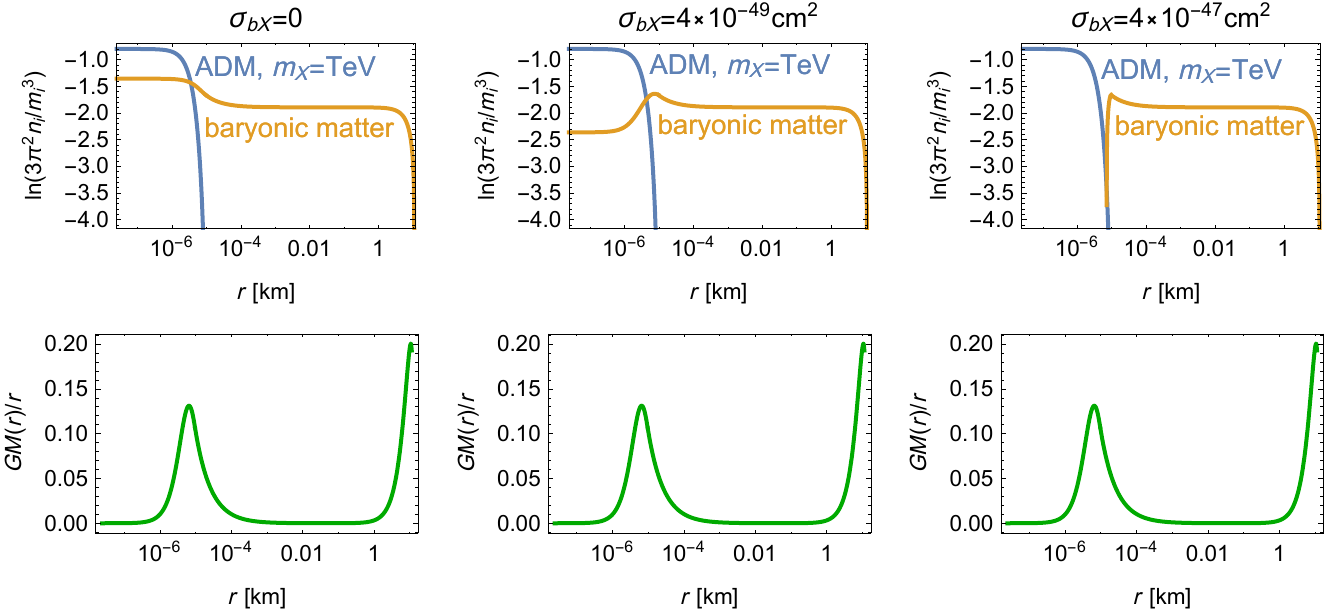}
\caption{Baryon number density (orange), ADM number density (blue), and compactness, $G M(r) / r$ (green), as functions of star radius for fermionic ADM with negligible self-interactions and $m_X = 1 \TeV$ admixed with baryonic matter. Since the number densities are normalized by $m_i^3$ and $m_b \sim \GeV$, the ADM number density in most of the core at small radius is more than $10^9$ times greater than typical baryon number densities. The total star mass is $1.5 \Msolar$ and $N_X = 0.396 (\Mplanck/m_X)^3$. The left-hand plots assume no ADM-baryon interactions while the middle and right-hand plots assume a repulsive ADM-baryon interaction as described in Sec.~\ref{sec: baryon-adm interactions} such that $\sigma_{b X} = 4 \times 10^{-49} \text{cm}^{2}$ and $4 \times 10^{-47} \text{cm}^{2}$, respectively. The EoS used for baryonic matter is as in Fig.~\ref{fig: free ADM admixed NS contour plot}.}\label{fig: admixed star profiles}
\end{figure*}

The left-hand plot in Fig.~\ref{fig: free ADM admixed NS contour plot} shows contours of constant $N_X$, $N_b$, and $M$ as functions of central ($r=0$) ADM and baryon number densities for solutions to the admixed star equations, Eqs.~\ref{chem potential} and \ref{metric TOV}, with non-interacting zero-temperature, $m_X= 100 \, m_b$ ADM, where $m_b\equiv939.5 \MeV$ is the nucleon mass scale. 
The baryonic matter was modeled with the HB EoS as described in \cite{Hinderer:2009ca} and shown in Fig.~\ref{fig: mX varies equilibrium sequences}, though this detail is unimportant. The important points are: 
\begin{itemize}
\item The constant $N_X$ contours are independent of central baryon density, and ${N_X}_\text{max}$ at fixed baryon number is the same as for an ADM-only star, demonstrating that baryonic matter affects neither the structure of the ADM core nor its stability endpoint.
\item The maximum $M$ and $N_b$ at any fixed value of $N_X$ are the same as a baryons-only star up to negligible fractions, demonstrating that the NS matter stability endpoint is unaffected by the ADM. This is because the ADM's contribution to the total mass of the NS is small: $m_X N_X \lll \Msolar$. The curvature of the $M$, $N_b$ contours indicates that the ADM affects the baryon density profile at the center of the star. C.f.~Fig.~\ref{fig: admixed star profiles}.
\item The $X$ number density and Fermi momentum for solutions near the stability endpoints, where $N_X = ({{N_X}|_{N_b}})_\text{max}$, are order $10^{-2} m_X^3$ and $m_X$, respectively---much greater than the baryon density, and also relativistic so that even if the NS is relatively warm, the zero-temperature approximation used for the ADM EoS is still valid.  
\end{itemize}
We checked that these features also hold for self-interacting ADM when $M^\text{ADM-only star}_\text{max} \ll \Msolar$. Thus for $f \ll 1$, \Eq{m collapse} is accurate. 

By contrast, the right-hand plot of Fig.~\ref{fig: free ADM admixed NS contour plot} shows similar contours when $m_X= m_b$. The shape and overlap of $M$, $N_b$, and $N_X$ contours along the diagonal from near the bottom left to top right are highly interdependent because here the ADM and baryonic matter densities are similar and both components contribute similarly to the total star mass.  For $N_b$ fixed at its value corresponding to a 1.5$\Msolar$ non-admixed NS ($N_b = 0.66 {N_b}_\text{max}$), the maximum-mass stable admixed star corresponds to $M \sim 1.7 \Msolar$ and $N_X \sim 0.4 {N_X}_\text{max}$. The amount of $m_X=m_b$ ADM that destabilizes the NS is smaller than the amount that destabilizes an ADM-only star. But the amount \emph{by mass} is a sizable fraction of the remainder of baryonic matter that would destabilize a baryons-only NS, and by \Eq{MDM captured} far exceeds the amount that can be captured. 

The main message conveyed by Fig.~\ref{fig: free ADM admixed NS contour plot} is: small amounts of fermionic ADM by mass ($m_X N_X \ll \Msolar$) cannot induce collapse of a NS unless the ADM is concentrated in a very dense ($\epsilon_X \gg \GeV^4$) core. Due to Fermi degeneracy pressure, this can occur only when $m_X \gg \GeV$. And in this case, the amount of ADM that destabilizes the NS is negligibly modified from the amount that destabilizes an ADM-only star.

We find that dense ADM cores can dramatically affect the density profile of baryonic matter overlapping the core, even though as noted above the maximum mass, radius, and baryon number at fixed ADM number is affected by negligible fractions. This is shown in Fig.~\ref{fig: admixed star profiles}, with baryon-ADM interactions as modeled in Sec.~\ref{sec: baryon-adm interactions} present or not. With even moderate $\sigma_{b X}$ well below direct detection constraints on large $m_X$ ADM, there are classes of solutions with vanishing $n_b$ at $r=0$---because of the high ADM densities, the repulsive interaction between ADM and baryonic matter wins over gravitational attraction and expels the baryonic matter from most of the core. In these cases we scan solutions by setting the ADM density at $r=0$ and the baryon density at the edge of the ADM core. We checked that including the baryon-ADM interaction at levels allowed by direct detection does not affect the conclusions described in the previous paragraphs. 

Finally we note that it could be interesting to examine the structure of NSs admixed with sub-GeV constituent mass ADM given repulsive ADM-baryon interactions---the ADM could be concentrated at the outer edge of the NS and beyond, which could lead to larger-than-naively-expected effects on the tidal deformability and/or moment of inertia of the NS for a given total amount of collected ADM.

\bibliography{ADMNuggets.bib}

\end{document}